\documentclass[journal,twoside]{IEEEtran}
\IEEEoverridecommandlockouts
\usepackage{todonotes}
\usepackage[hyphens]{url}

\title{Blockchain Mutability: Challenges and Proposed Solutions}

\author{Eugenia Politou and Fran Casino, \IEEEmembership{Member, IEEE} and Efthimios Alepis and Constantinos Patsakis, \IEEEmembership{Member, IEEE}
\thanks{E. Politou, F. Casino. E. Alepis and C. Patsakis are with the Department of Informatics, University of Piraeus, Greece. }}

\usepackage{graphicx}

\begin{document}

\maketitle

\begin{abstract}
Blockchain's evolution during the past decade is astonishing: from bitcoin to over 2.000 altcoins, and from decentralised electronic payments to transactions programmable by smart contracts and complex tokens governed by decentralised organisations. While the new generation of blockchain applications is still evolving, blockchain's technical characteristics are also advancing. Yet, immutability, a hitherto indisputable property according to which blockchain data cannot be edited nor deleted, remains the cornerstone of blockchain's security. Nevertheless, blockchain's immutability is being called into question lately in the light of the new erasing requirements imposed by the GDPR's ``\textit{Right to be Forgotten (RtbF)}''  provision. As the RtbF obliges blockchain data to be editable in order restricted content redactions, modifications or deletions to be applied when requested, blockchains compliance with the regulation is indeed challenging, if not impracticable. Towards resolving this contradiction, various methods and techniques for mutable blockchains have been proposed in an effort to satisfy regulatory erasing requirements while preserving blockchains' security. To this end, this work aims to provide a comprehensive review on the state-of-the-art research approaches, technical workarounds and advanced cryptographic techniques that have been put forward to resolve this conflict and to discuss their potentials, constraints and limitations when applied in the wild to either permissioned or permissionless blockchains.
\end{abstract}

\section{Introduction}
Blockchain technology dominates today's news, discussions, and articles, whereas its initiatives proliferate across industry and academia. Yet, few technologies today are as misunderstood as blockchain. For some, blockchain is just a hype, an immature solution \cite{bitcoinexchange2018}, an exaggerated bubble \cite{gerard2018blockchainidentity}, or even a crypto-medieval system \cite{medium2018blockchaincruppy}. For others, it is an undeniably ingenious invention, an advance, a revolutionary technology. Blockchain's technological breakthrough has been even compared to the one brought by the use of the TCP/IP to modern computing or the one Linux brought to modern application development \cite{lakhani2017truth,swan2015blockchain,8633601,zheng2016blockchain,mearian2019whatisblockchain}. In addition, the bitcoin, the first cryptocurrency that exploits blockchains, has been called as ``digital gold'' \cite{lunobitcoingold}, while ethereum, the largest open-source blockchain-based distributed computing platform, has been characterised as the backbone of the new Internet \cite{ethereumbackbone}.

Even though its underlying technology existed long before Satoshi Nakamoto published his paper on bitcoin \cite{nakamoto2008bitcoin}, the immense and profound impact the bitcoin had in financial trades worldwide revealed blockchain as a new highly promising direction for decentralised computing. In the wake of the 2008 financial crisis where consumers' trust in banking was shaken, bitcoin's notion of decentralised financial systems seemed particularly appealing. Nevertheless, while blockchain technology is commonly associated with bitcoin and other cryptocurrencies, these are just the forerunners of a whole new wave of blockchain applications. According to experts, apart from disrupting financial services, blockchain could end up transforming a number of important industries, from healthcare to politics, whereas has the potential to create new foundations for economic and social systems \cite{casino2018systematic,swan2015blockchain,panarello2018blockchain}. As most of its broad possible applications are still emerging, the future orientation and impact of blockchain technology cannot be easily predicted. Still, its first stages of development during its decadal lifetime are beyond any expectations.

Undoubtedly, blockchain's substantial impact on current and future real-world applications is attributed to its most profound quality, its trustlessness. Trustlessness stems from blockchains' inherent security and transparency which eliminate the need for a third party intermediation and trust among users in decentralised and untrusted environments \cite{marsh2005trust}. A fundamental property that underpins the blockchain's secure and transparent nature, and therefore guarantees its transactional integrity and auditability, is immutability. Blockchain's immutability certifies that transaction data residing in blockchains are tampered-proof, i.e. they can neither be removed nor mutated. However, this append-only data structure signifies the permanent storage and availability of the stored information to everyone in the blockchain network. Clearly, this property, albeit desirable in some contexts, contradicts several privacy requirements and data protection rights when personal data are at stake. Among others, it clearly challenges the Right to be Forgotten (RtbF) defined in the new European data protection regulation, the GDPR, according to which individuals have the right to delete their personal data if certain conditions apply \cite{gdpr}.

Acknowledging the above contradiction, considerable research is carried out nowadays to design and develop methods for allowing the modification or deletion of blockchain data while maintaining its security, auditability and transparency. As the conflict around blockchain's immutability may affect the adoption of blockchains substantially to a broad area of applications, we believe that resolving such disputed areas will be to the advantage of both academia and industry. To that end, this paper aims to provide a comprehensive review on the state-of-the-art research approaches, technical workarounds and advanced cryptographic techniques that have been put forward to resolve this conflict and to discuss their potentials, constraints and limitations when applied in the wild to either permissioned or permissionless blockchain settings.

The rest of the work is structured as follows. Since the heart of blockchain lies in the decentralisation, in the following section we describe the decentralised architecture in terms of blockchain technology. Next, we present blockchain's key characteristics relevant to our work, namely permissions, consensus protocols, trustlessness, privacy, transparency, and most importantly, immutability. In section 4, we discuss the collision of blockchain's immutability with the GDPR's RtbF, whereas in section 5 we review the currently employed technical methods and the state-of-the-art techniques introduced to comply blockchains with the erasing requirements of the RtbF. The paper concludes by discussing the tension around blockchain's evolution and the respective challenges in terms of its alignment with the RtbF.

\section{Decentralized architectures}
Although the hype of decentralisation has been demonstrated during the late years by the boom of Distributed Ledger Technologies (DLTs), decentralisation of information systems is not a new idea. Even from the early 70's distributed and decentralised architectures were introduced to eliminate the problem of single point of failure and to increase systems' robustness. It is worth pointing out, however, that while the terms decentralized and distributed are commonly used interchangeably to denote the lack of a central point of control, they actually have a subtle different meaning; the former is used to describe the conceptual and logical model of control, while the latter describes the technical characteristics of the infrastructure used to be built upon \cite{decentralizedvsdistributed}.

Since the dawn of online social networking, decentralisation has also been proposed as an alternative for enhanced privacy and personal sovereignty in online social context \cite{yeung2009decentralization}. In recent years, decentralisation has been re-introduced as a mean to assure the reliability of non-trusted environments such as those of electronic currencies, i.e. cryptocurrencies. Nowadays, cryptocurrencies are usually discussed in the context of blockchains and distributed ledger technologies, terms closely interrelated but not identical. In what follows, we summarise and clarify the notions of DLT, blockchain and cryptocurrency and highlight their respective differences.

\subsection{DLT}
A DLT is a distributed digital ledger stored on a network of machines. Any changes to the ledger are reflected simultaneously for all holders of the ledger while the information stored is authenticated by a cryptographic signature \cite{deshpande2017distributed}. The decentralized nature of the DLT eliminates the need for a central authority or intermediary to process, validate or authenticate transactions. At their core, DLTs are data structures to record transactions and set of functions to manipulate them. While each DLT differentiates itself using different data model and technologies, generally all DLTs are based on three well-known technologies: public key cryptography; distributed peer-to-peer networks; and consensus mechanisms. All three are blended in a unique and novel way to operate in an untrusted decentralised environment \cite{el2018review}.

\subsection{Blockchains}
Even though blockchain technology was first outlined in 1991 as an effort to implement a system where document timestamps could not be tampered with \cite{haber1990time}, it was not until January 2009 that blockchain attracted worldwide attention when its first real-world application, the bitcoin cryptocurrency, was launched \cite{nakamoto2008bitcoin}. Although the terms DLT and blockchain are often used interchangeably in the literature, they are not equivalent. For instance, while a blockchain is a sequence of blocks, DLTs do not require such a chain. As a matter of fact, a blockchain is just one type of DLT formed by a linked list (chain) of blocks connected to each other using hash codes, where each block references the previous block in the chain. Each block may contain a series of transactions which can be data of any sort. In blockchains, the transaction data are continuously appended, and they can be accessed by all the network participants (nodes). Essentially, blockchains are distributed and immutable ledgers that store transactions history while they provide a set of features that differentiate them from the other DLTs: smart contracts, which are pieces of executable code residing on the blockchain and executed once specific conditions are met; and miners, which are mining new transactions into the blockchain and can benefit financially from these mining activities \cite{el2018review}.

\subsection{Cryptocurrencies}
While there have been multiple attempts during the last 30 years to solve the complex issues surrounding digital currencies \cite{cypherpunksandtherise,untoldstoryofbitcoin,howtomakeamint}, this was not achieved before 2009 when the bitcoin was launched. Generally speaking, the term cryptocurrency refers to a decentralised cryptography-based currency. Cryptocurrencies can be seen as asset resources or tokens on a blockchain network, and they are just one of the many possible applications of blockchain. Arguably, the true value of blockchain technology goes far beyond cryptocurrencies, whereas a blockchain can stand on its own just fine - no cryptocurrency needed \cite{christidis2016blockchains}. In fact, there are already blockchain frameworks without any built-in cryptocurrency \cite{endingthebitcoin}. Yet, cryptocurrencies currently underlie most of the public blockchain applications to facilitate and incentivise their transactions.

Although bitcoin is currently the dominant cryptocurrency used in decentralised payments, the number of alternative cryptocurrencies (altcoins) has already surpassed 2.000 \cite{coinmarketcap}. In the context of cryptocurrencies like bitcoin, the name refers to more than the underlying technology since it can also be used to denote the protocol, the software system that transfers the money over the blockchain ledger, as well as the token, i.e. the currency itself that is traded in transactions or exchanges \cite{swan2015blockchain}. Nevertheless, while referring to the token as the technology can be right in the case of bitcoin, this is not the case when dealing with other blockchain projects like ethereum \cite{buterin2014ethereum} where the technology is known as ethereum, the native token is ether, and transactions are paid in gas.

\section{Blockchain characteristics }
Over the last few years, blockchain is rapidly moving from the fat protocols stage, where all value is generated in the protocol layer, into the fat Decentralised Applications (DApps) stage where transactions are programmable by smart contracts, and complex tokens are governed by Decentralised Autonomous Organisations (DAOs) \cite{swan2015blockchain,nomorehype}. As this new generation of applications is evolving, blockchain's technical characteristics and specifications are becoming even more advanced and sophisticated \cite{casino2018systematic}. By all means, describing in detail all the blockchain's features and functions is beyond the scope of this paper. Instead, for the sake of simplicity, we delve into blockchain's characteristics that are relevant to our following discussion regarding blockchain's immutability and its collision with the RtbF.

\subsection{Permissionless and permissioned blockchains}
Typically, there are two types of blockchains: permissionless and permissioned. A permissionless blockchain is considered a public one in the sense that anyone can be a node and interact with the network by either submitting transactions, and hence adding entries to the ledger, or participating in the process of transaction verification and block mining, or even creating smart contracts. In other words, anyone can read the chain and write a new block into the chain. In contrast, permissioned blockchains limit the parties who can transact on the blockchain and can contribute to its state. Actually, in a permissioned blockchain, only a restricted set of users have the rights to see the recorded history, to validate the block transactions, to issue transactions of their own, or to create smart contracts. Permissioned blockchains can be either private or consortium blockchains. A private blockchain is fully controlled by an organisation and only nodes from this specific organisation could determine the final consensus. Several organisations construct a consortium blockchain and only a group of pre-selected nodes are responsible for validating the blocks, and thus for participating in its consensus process \cite{meiklejohn2018top}. Apparently, permissioned blockchains, acting as closed ecosystems in which some central authorities control participation, cannot be regarded as fully decentralised networks since a minimum level of trust among the nodes is sustained. Instead, consortium blockchains are regarded as partially decentralised, while private blockchains have been compared to centralised networks, and even to distributed databases \cite{howarebetterthandatabases}.

Permissionless blockchains usually employ fat protocols that compensate network contributors with tokens. On the other hand, permissioned blockchains generally do not need to employ a cryptocurrency model or monetary tokens due to the nature of these business networks. Nonetheless, both types of blockchains have their own advantages and disadvantages and can be suitable for different kind of situations. Although it may seem that in an institutional context private blockchains is unquestionably a better choice, it has been argued that public blockchains operating within or across organisations still have a lot to offer \cite{onpublicandprivate}.

\subsection{Maintaining trust through consensus protocols}

In view of the fact that a third party is no longer needed in a blockchain to verify data integrity and to maintain trust, as opposed to the centralised architectures, consensus algorithms are used to maintain data consistency \cite{meiklejohn2018top}. To put it another way, a consensus protocol allows all nodes of the blockchain, and the DLTs in general, to agree on a single version of the truth, i.e. on the transactions and the order in which these are listed on the newly-mined block, without the need of a trusted third party. Otherwise, the individual copies of the ledger will diverge and it will end up with branches, called forks, of the chain; the nodes will have a different view of the global state \cite{christidis2016blockchains}. As previously mentioned, while every node in a permissionless blockchain could take part in the consensus process, only a selected set of nodes are responsible for validating the block in a permissioned blockchain. Some of the main consensus protocols used as of today are Proof of Work (PoW), Proof of Stake (PoS), Delegated Proof-of-Stake (DPoS), Proof of Authority (PoA), and Practical Byzantine Fault Tolerance (PBFT).

In PoW, which is the underlying consensus of the bitcoin, several nodes of the distributed ledger, called miners, compete to solve a complicated mathematical problem, that is to calculate a hash value of the block header equal to or smaller than a threshold, and hence to validate a block of transactions. Once the first miner finds a solution, it broadcast it to the other nodes which then verify the solution by mutually confirming the correctness of the hash value. If all the nodes agree on the solution, the consensus is reached, and the new block is appended to all the ledgers held by the nodes of the network. The idea is that the solution to the problem is hard to find but easy to verify by the rest of the network. While there might be cases of multiple nodes finding a solution nearly at the same time, and hence valid blocks to be generated simultaneously resulting thereby in forks, these cases are extremely unlikely, albeit not impossible \cite{thestoryofdao}. Nevertheless, a chain that becomes longer thereafter is judged as the authentic one. The PoW is characterised by its high energy consumption, since a huge amount of computational power is required for solving the mathematical puzzle to mine a block. Moreover, in PoW there is always the possibility of the formation of mining pools, i.e. groups of miners who pool their resources together and potentially could control the network. In PoS, which is regarded as an energy-saving alternative to PoW, miners have to prove the ownership of the amount of currency since it is believed that people with more currencies would be less likely to attack the network \cite{meiklejohn2018top}. DPoS is a more efficient PoS mechanism that uses a reputation system and real-time voting to achieve consensus. Nodes vote for representatives to secure their network and representatives are rewarded by validating transactions for the next block. In PoA, transactions are validated by approved accounts, known as validators. By attaching a reputation to an identity, validators are incentivised to uphold the transaction process, to avoid having their identities linked to a negative reputation. PBFT reach a consensus without the energy consumption required by PoW. The consensus decision is determined based on the total decisions submitted by all the nodes and the honest nodes come to an agreement of the state of the system through a majority.

By definition, consensus protocols in permissionless blockchains promote and establish decentralised trust in non trusted environments. This is the result of their employed incentive mechanisms which rely mostly on game-theoretic principles for the correct operation and assume absolute non-trust among the participants. Instead, the consensus protocols operate on the assumption that all miners behave in a way that is profitable to them \cite{thurimella2018hitchhiker}. In an ideal scenario where there would be a minimum level of trust, all validating nodes would vote on the order of transactions for the next block, and they would go with what the majority decides \cite{christidis2016blockchains}. However, due to the complete absence of trust in permissionless blockchains, nodes cannot rely on each other, and therefore, they are rewarded with incentives for correct behaviour to collectively agree on the state of the ledger \cite{thurimella2018hitchhiker}. In PoW for instance, if a malicious user tries to subvert the system by creating a fork and entering into a race with other miners to create an alternate ledger, the resulting computational cost will be tremendous, even in the case of winning. Instead, if the same work is directed towards honest mining, it can possibly result in bigger profits by way of incentives. Hence, trying to defraud the system is generally not in one's interest. This is why the trustlessness of PoW consensus mechanism, which makes no assumptions about the honesty or reliability of participants, is currently considered more suitable for permissionless blockchains. On the contrary, due to the risk of Sybil attacks public blockchains cannot rely on the PBFT consensus algorithm which requires a majority of honest nodes: even when there is only one malicious participant, it can create multiple fake identities, get multiple votes, and thus influence the network to favour its interests, forcing the number of honest nodes to a minority \cite{thurimella2018hitchhiker,christidis2016blockchains}.

\begin{figure}[th]
    \centering
    \includegraphics[width=\columnwidth]{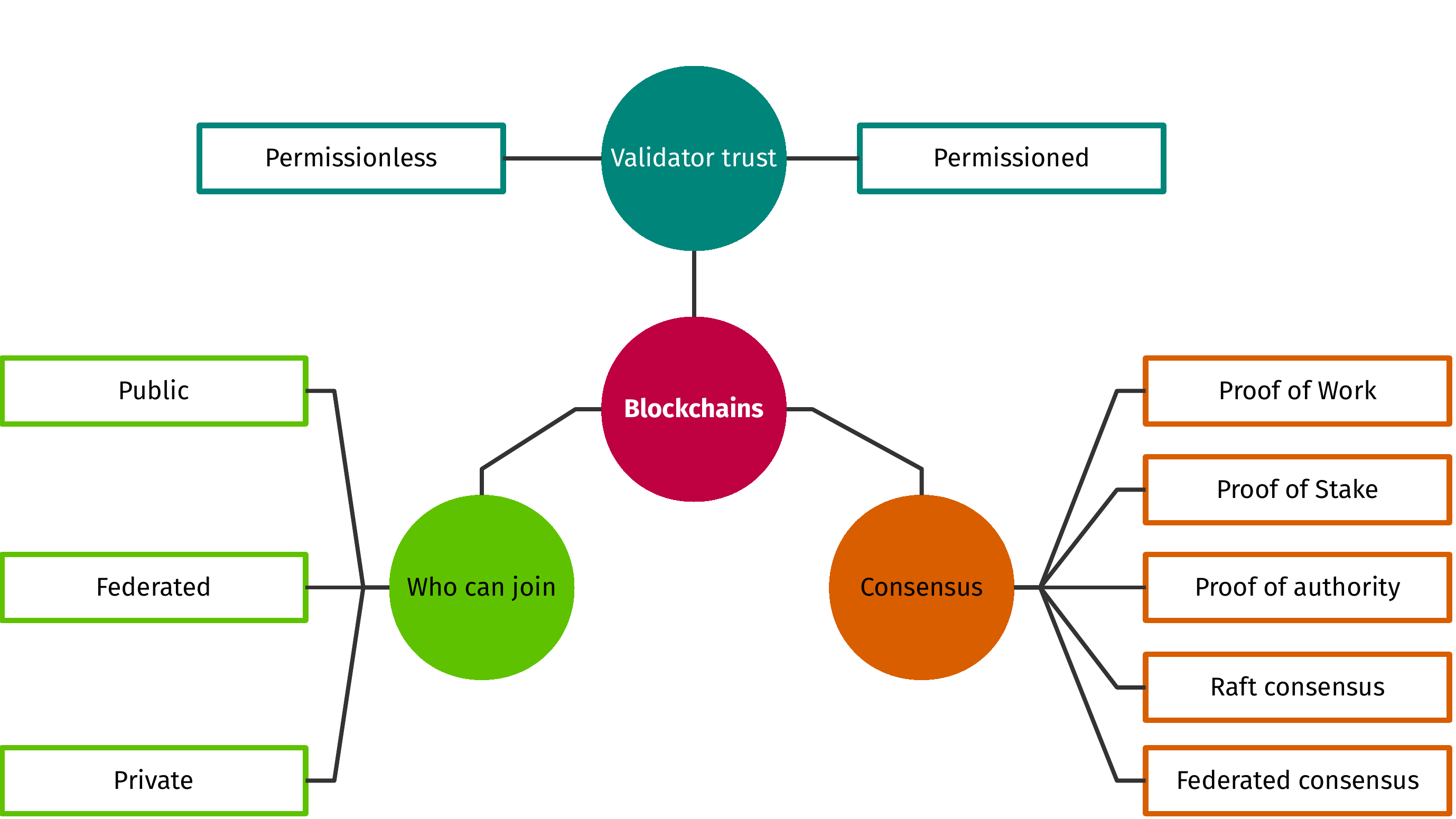}
    \caption{Blockchain categorisation.}
    \label{fig:categorisation}
\end{figure}

The consensus mechanisms employed in permissioned blockchains can be the same as in permissionless networks or can be completely uniquely designed (e.g. authority-based). In fact, it has been argued that consensus based on cryptocurrency is unsustainable for enterprise use and permissioned blockchains \cite{prosandconsofdifferent}. For instance, Hyperledger Fabric \cite{hyperledger}, a permissioned blockchain infrastructure oriented towards enterprises, does not require a built-in cryptocurrency because consensus is not reached via mining \cite{usingblockchain}. Generally speaking, given the trusted model of permissioned blockchains and the known identities of the network participants, the consensus mechanisms used are computationally inexpensive when compared to PoW as there is no need for protection through mining. In fact, private blockchains are far less costly to operate since, as long as the majority of validators are following the rules, blocks only need a simple digital signature from the nodes that approve them instead of expensive consensus protocols \cite{greenspan2017blockchain}. Most common voting-based consensus protocols preferred by permissioned blockchains are based on the family protocols of Paxos and PBFT \cite{thurimella2018hitchhiker,lamport2001paxos,cachin2017blockchain}. Given the extensive length of consensuses used in DLTs and the peculiarities of each one, only the main consensus protocols employed in blockchains were briefly mentioned here. The interested reader may further refer to \cite{meiklejohn2018top,cachin2017blockchain,sankar2017survey}.  A brief categorisation of blockchains based on some of their basic characteristics is illustrated in Figure \ref{fig:categorisation}.

\subsection{Privacy and transparency}
By design, blockchains are based on the principle of complete transparency according to which transactions, even if they hashed or encrypted, are visible to all participating nodes so that they can be validated \cite{christidis2016blockchains}. Therefore, since the content of every transaction is exposed to every node on the network, transactional privacy in blockchains is hard to be attained. Nevertheless, in permissioned blockchains where the nodes are known, privacy and confidentiality are usually preserved much more efficiently than in permissionless settings through the use of access control policies. On the other hand, while user accounts in permissionless blockchains can largely stay anonymous, and as such are thought to provide a series of privacy benefits to their users, many studies have demonstrated that there are still considerable risks to users' privacy \cite{meiklejohn2013fistful,goldfeder2018cookie,meiklejohn2018top,primavera2016interplay,biryukov2014deanonymisation}. For instance, research has shown that even when users are hiding behind multiple pseudonyms, these can be correlated and often identify them \cite{tschorsch2016bitcoin,meiklejohn2013fistful,fleder2015bitcoin,androulaki2013evaluating}. Adding to this the fact that transactions are linked, one can retrieve the full history of all transactions performed on a blockchain \cite{tschorsch2016bitcoin}.

Due to the transparent and permanent nature of blockchain technology which obliges data to be stored forever and to be publicly available to the entire network, putting personal data on blockchains has been broadly discouraged. As it has been argued, storing personal data into blockchains it is like having again ``Cambridge Analytica'' - a severe surveillance scandal - but on the blockchain \cite{gerard2018blockchainidentity}. Nevertheless, blockchains do not have to expose personal data directly to reveal individuals' personal information. By exploiting metadata information and by applying big data analytics potentially sensitive information can also be retrieved, e.g. recording visits to health practitioners may reveal sensitive details on someone's health status \cite{primavera2016interplay}. As it has been demonstrated in the literature, achieving privacy in a lightweight and flexible manner for all DLTs, in general, is still an open research question \cite{meiklejohn2018top}. That being said, it is worth noting that privacy was never one of blockchain's original problems to be addressed. As Buterin, the founder of the ethereum blockchain puts it, ``\textit{blockchains do not solve privacy issues, and are an authenticity solution only}'' \cite{privacyontheblockchain}.

In spite of this limitation, several approaches based on cryptographic techniques such as homomorphic encryption, zero-knowledge proofs \cite{hopwood2016zcash} and secure Multi Party Computation (MPC) \cite{zyskind2015enigma} have been proposed to address transactional privacy in blockchains. Broadly speaking, these techniques enable specific computations to be performed without revealing the inputs and outputs of those computations. These methods, however, are resource intensive so it is almost impossible to be implemented at scale \cite{christidis2016blockchains}. Tumblers or mixing services have also been used intensively lately as a mean to provide strong notions of anonymity in public blockchain networks \cite{meiklejohn2018mobius}.

\subsection{Blockchain immutability}
Immutability, or irreversibility, is a fundamental blockchain property that stems from the fact that transactions cannot be edited or deleted once they are successfully verified and recorded into the blockchain. This property is the consequence of the cryptographically linked blocks which are chained together with the hash value of the preceding (parent) block. In particular, each block contains a reference to the preceding block by including in its header a cryptographic hash of the transaction data within the preceding block. This cryptographic hash is actually calculated using a Merkle tree on all the transactions of the block. A Merkle tree is a data structure constructed by recursively hashing pairs of transactions until there is only one hash, called the Merkle root. Merkle trees are used in bitcoin to summarize all the transactions in a block, producing an overall digital fingerprint of the entire set of transactions, providing thus a very efficient process to verify whether a transaction is included in a block without the need for a complete local copy of all transactions. Since the root is known and secured through the mining process, branches can be loaded on demand from untrusted sources. The cryptographic hash algorithm used in bitcoin’s merkle trees is SHA256 applied twice, also known as double-SHA256. In bitcoin blockchains, simplified payment verification (SPV) based on Merkle tree is used in order to keep the size and the computational effort low, whereas in ethereum a variation of Patricia Merkle Tree is used.

In simpler terms, the Merkle root which comprises the information from all transactions of a given block, is included in the block header of the subsequent (child) block. Bearing in mind the collision-resistant property of the hash functions, any change of the transaction data in a block will change the hash of this block and, to maintain the integrity of the parent-child reference, will necessitate a change in its reference within its child block. This cascade effect ensures that once a block has many generations following it, it cannot be changed without forcing a recalculation of all subsequent blocks since such a recalculation would require enormous computation \cite{antonopoulos2014mastering} for Proof of Work-based protocols. The longest the chain of blocks a blockchain has, the more resilient the blockchain is to data tampering attacks because if an adversary modifies data anywhere in the blockchain, it will result in the hash pointer in the subsequent block being incorrect \cite{narayanan2016bitcoin,tschorsch2016bitcoin}.  Therefore, for properly deployed blockchains, data residing in blockchains cannot be ever mutated or removed. Even though tampering with data already stored in the blockchain is not possible, data can be appended to the blockchains. Therefore, blockchains are known as append-only, tampered-proof and immutable data structures. Inevitably, since blockchain's immutability assures its transactional integrity, i.e. the correct and permanent storage of blocks and transactions within the blockchain, it is of paramount importance to blockchain's security and a cornerstone of its highly praised values of trustlessness and censorship-resistance.

While, as demonstrated, it is impossible to delete, update or rollback transactions once they are included in a blockchain, some would argue otherwise: considering that immutability is an emergent, and not intrinsic, property of a blockchain data structure, and therefore an agent or set of agents with a sufficient amount of computing power can modify it, stating that a blockchain is by default immutable is incorrect and misleading \cite{conte2017blockchain,finck2018blockchains,8653269}. Especially in the context of permissioned blockchains where the number of nodes is limited, tampering with blockchain data should not be regarded as impossible since there is always a possibility of the majority of the consortium or the dominant organisation nodes to vote for their version of truth and to amend the ledger accordingly \cite{meiklejohn2018top,swanson2015consensus}. Hence, although in public blockchains the existence of a long chain of blocks makes the blockchain's deep history immutable due to the extremely high cost involved for altering the hash-based integrity of the blocks, ensuring immutability in private blockchains is much cheaper and stronger, as long as the majority of validating nodes are following the rules \cite{greenspan2017blockchain,antonopoulos2014mastering}. However, it has been argued that even in public blockchains there is no such thing as perfect immutability since under certain conditions a particular blockchain can be changed \cite{greenspan2017blockchain}. Although events such as the Ethereum Decentralized Autonomous Organization (DAO) fork clearly align with such claims \cite{understandingthedaohack}, these hard forks are exceptionally rare and definitely cannot be applied on a regular basis. Hence, it is commonly held that altering transactional data in public blockchains is thus far practically impossible.

\section{Blockchain immutability and the Right to be Forgotten}
Blockchains by definition are unable to forget since tampering with transactional data stored in blockchains has been identified as nearly impossible \cite{finck2018blockchains}. Indubitably, this immutable and transparent record keeping of blockchain data facilitates the movement and storage of information in a secure, auditable and credible way, and consequently guarantees blockchains' credibility, persistency and security. Despite its apparent benefits, blockchain immutability also has some unintended consequences such as when erroneous or illegal content is stored in the blockchain \cite{thurimella2018hitchhiker}. Likewise, as already discussed in Section 3.3, blockchain immutability presents several risks to people's privacy. More precisely, immutability's collision with privacy and data protection rights renders absolute immutability a major barrier to blockchain's adoption when personal data are at stake \cite{schwerin2018blockchain}. In this regard, immutability, a hitherto indisputable property and the cornerstone of blockchain's security, is being called into question in the light of the erasing requirements imposed from the recently adopted European data protection regulation, the GDPR. Although the GDPR provides strict requirements for the processing of the personal data and offers extended legal rights to individuals residing in the EU, in its provided recitals and articles it does not take into account decentralised technologies such as DLTs and blockchains. On the one hand, this was because regulators deliberately chose to follow a technology-agnostic approach in order not to bind the provisions of the law with current trends and state-of-the-art technologies in computer science \cite{politou2018forgetting}. On the other hand, however, this was because over the long period under which the final GDPR text was being debated and finalised, blockchain technology has not been a widespread technological trend that is these days. As a result, various legal and technical divergences and incompatibilities between the GDPR and the blockchain technology have been unavoidably identified \cite{finck2018blockchains,schwerin2018blockchain,sater2017blockchain,zetzsche2018distributed,ibanez2018blockchains}.

Of the GDPR's provisions, the most profound and controversial one is the Article 17 that anticipates the Right to be Forgotten (RtbF), namely the possibility of individuals to request the erasure of their personal data when certain conditions are met (Article 17(1)). In particular, the RtbF entails the permanent deletion of personal data upon request and from all the places to which they have been disseminated \cite{politou2018forgetting}. As already thoroughly discussed and analysed in previous works, the impact of encompassing the RtbF on contemporary information systems is immense, whereas its integration into the design of future technological developments is currently disputable \cite{politou2018forgetting,politou2018backups}. Blockchain technology, due to its immutability, is one such advanced development that contradicts the RtbF. Although one might argue that anonymizing personal data residing in blockchains through public key cryptography is a reasonable step for blockchain data to fall outside of the scope of GDPR, it should be outlined that private and public keys as well as hashed data are pseudonymous, not anonymous, and therefore also qualify as personal data under the GDPR (since pseudonymous data are still personal and consequently they are not exempted from the Regulation (Article 4(1))) \cite{wp29consent,politou2018forgetting,finck2018blockchains}. Put differently, blockchain compliance with the GDPR only through the use of hash values and public key cryptography cannot be guaranteed \cite{schwerin2018blockchain}. Taking further into account that data stored in blockchains are never completely anonymous (Section 3.3), it is apparent that the RtbF strikes at the heart of the blockchain's immutability property.

Against this background, CNIL, the French Data Protection Authority, notes that it is technically impossible to grant the data subject's request for erasure when data is entered in the blockchain. In fact, while the CNIL recognises that there are some cryptographic methods that may make the data ``almost inaccessible'', it still questions the extent to which these solutions provide full compliance with the GDPR since the solutions do not ``\textit{strictly speaking, result in an erasure of the data insofar as the data would still exist in the blockchain}'' \cite{martin2018blockchain,cnilblockchain}. Along the same lines, the European Data Protection Supervisor (EDPS) stresses the importance of enabling the manageability of the personal data, i.e. their alteration, deletion, and selective disclosure, as a mean to maintain people's privacy \cite{edpsopinion2018}. Furthermore, a recent resolution from the European Parliament on DLTs and blockchains raises the need for blockchain applications to be compliant with the GDPR, stressing the fact that the RtbF is not easily applicable to this technology \cite{parliamentresolution2018}.

Inevitably, the RtbF has been seen by many blockchain advocates and crypto activists as an obstacle for expanding the blockchain technology to a broad area of applications. Still, others have argued that approaches for adding preapproved, limited, and transparent methods to alter data on an immutable system is a trade-off necessary to be able to utilise the advantages of the blockchain technology \cite{schwerin2018blockchain,sater2017blockchain}. In this respect, the World Economic Forum has sounded the alarm about the struggling of blockchain innovation due to the GDPR and urged for flexible policy frameworks to allow the benefits of data and technology to be realised \cite{weforum2018}. Ideally, for enabling data deletion, the participants of a blockchain would have to agree on an effective process to jointly execute a lawful request to erase personal data from the decentralised ledgers \cite{wirth2018privacy}. As already discussed, in permissioned blockchains where there are specific entities (authorities or enterprises) in charge and legally accountable, introducing mutability in the blockchain without interrupting its functionality should not be considered an impossible task \cite{swanson2015consensus,meiklejohn2018top}. In this perspective and in the context of permissioned blockchains, the term ``pragmatic immutability'' has been coined to pave the way for greater blockchain adoption outside the world of cryptocurrency \cite{accentureimmutability}.

However, introducing mutability in permissionless blockchains is rather challenging due to the absolute lack of trust among the participants. Yet, there exist optimistic voices that put their faith in advanced cryptographic techniques to guarantee individual privacy in decentralised architectures such as blockchains \cite{primavera2016interplay}. With this in mind, several research works have been carried out lately in an attempt to conform blockchains to the RtbF erasing requirements and to consequently adjust them to privacy-intensive applications. Among others, these works include technical workarounds and advanced cryptographic methods to either bypass or remove blockchain immutability both for permissioned and permissionless blockchains. The state-of-the-art of these works is discussed hereafter.

\section{Current efforts for balancing immutability and the RtbF}
\begin{figure*}[th]
    \centering
    \includegraphics[viewport=0bp 0bp 794bp 370bp,clip,width=.8\textwidth]{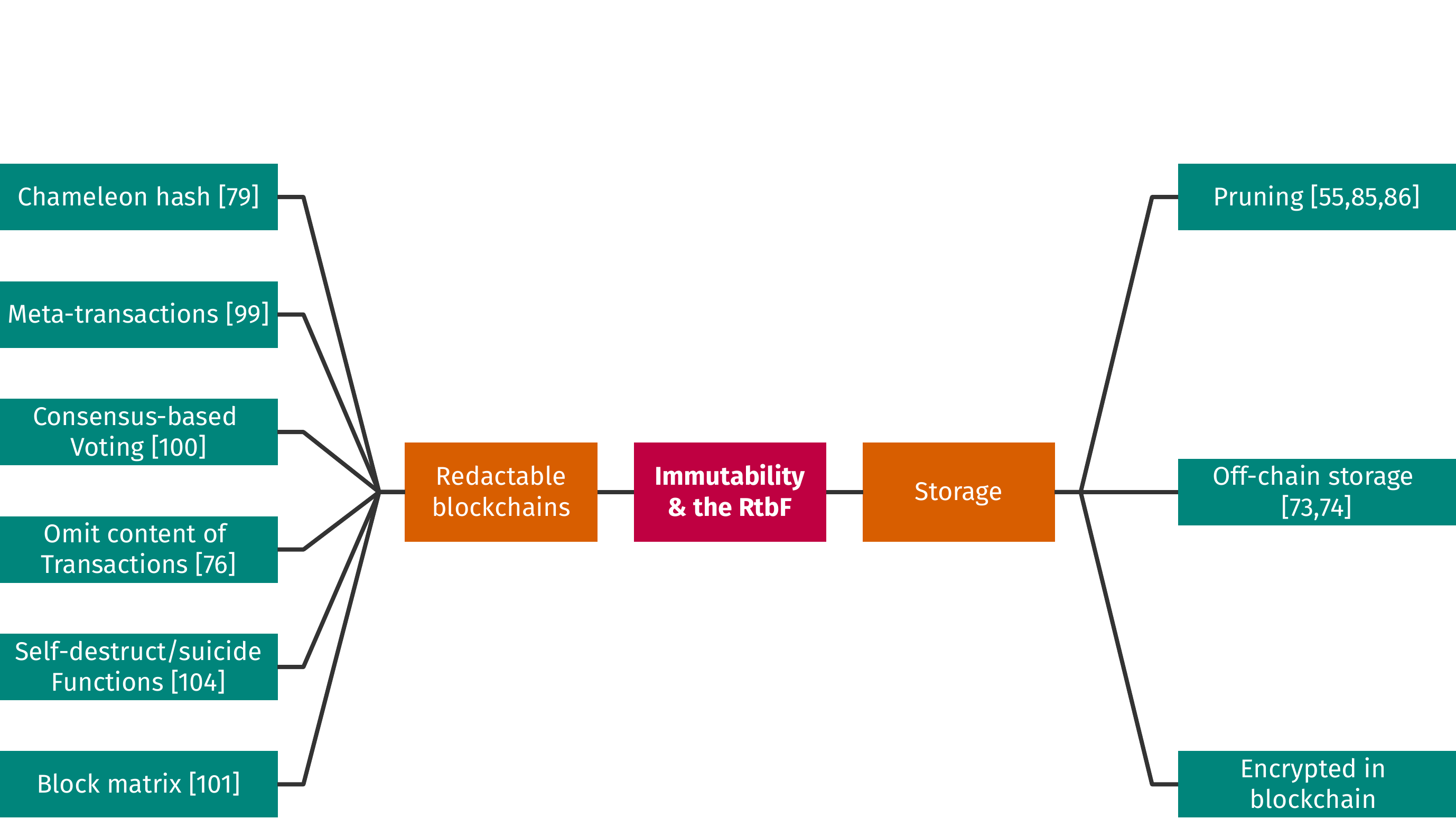}
    \caption{An overview of solutions for balancing immutability and the Right to be Forgotten.}
    \label{fig:sols}
\end{figure*}

To address privacy issues arising from blockchains, and particularly to tackle the controversy around blockchain's immutability and the RtbF, various approaches have been embraced by researchers and information technologists. These approaches focus on either circumventing or conditionally removing blockchain's immutability. An overview of these solutions is illustrated in Figure \ref{fig:sols}.

\subsection{Bypassing blockchain's immutability}
A common workaround suggested throughout the literature for aligning blockchains with the GDPR privacy requirements is the use of blockchains only for storing a timestamp and a hash that point to the actual information held off-chain \cite{eberhardt2017or,garcia2017deploying}. Therefore, when information needs to be amended or deleted only the fact that the specific content version existed at a given point in time will remain in the blockchain. Bearing in mind that by using the stored hash alone the original content cannot be reconstructed, this workaround seems to resolve the blockchain's immutability collision with the RtbF rather elegantly. Indeed, off-chaining techniques so far are considered to be key tools in blockchain-based application engineering as they present significant benefits, such as reduced blockchain data storage requirements, and hence fewer scalability issues, and GDPR compliance \cite{eberhardt2017or,finck2018blockchains}. As a matter of fact, a recently conducted study \cite{schwerin2018blockchain} among experts concluded that blockchains could be indeed compliant with the privacy by design principles of the GDPR, and consequently with the RtbF, by employing these kind of off-chaining techniques. On the downside, however, these techniques move the responsibility of robust, distributed data storage to other protocols like the IPFS \cite{ipfs}, while they introduce complexity and additional delays. Furthermore, they have been criticised for decreasing blockchains' security by introducing more attack vectors \cite{dorri2019mof,blockchaingdprparadox}. But most importantly, these solutions do not avoid the burden of having to remove the hash pointers from the blockchain since hashed data are pseudonymous, not anonymous, and therefore need to be protected \cite{wp29consent}. For instance, hashed data may reveal sensitive personal information either when combined with other available information or when they are subject to dictionary attacks.

Another alternative solution for complying blockchains with the RtbF is to have the data stored in the blockchain in an encrypted form, and when the user asks to delete personal information forgetting or deleting the encryption key will make the data inaccessible, i.e. no retrievable. Although some experts argue that in the case of the blockchain inaccessibility equals deletion, this is not the opinion of the data protection authorities such as the French CNIL which explained that, strictly speaking, this approach is not an actual erasure \cite{martin2018blockchain}. Another limitation of this solution stems from the difficulty in managing the decryption keys among many parties that need access to the data. Furthermore, there is always the case that personal data to become unreadable or available to everyone when the key is either lost or becomes accidentally known \cite{opendata2016,politou2018backups,ateniese2017redactable}. Taken into account that data shall remain encrypted across their life cycle, a further limitation derives from the rapid advancements in quantum computing which, according to experts, is going to break most encryption schemes used nowadays \cite{aggarwal2017quantum,ibmunveilsfirstquantum,quantumcomputers, blockchainimmutabilityblessing}. To avoid information be susceptible to decryption once quantum computers become available, sensitive data need to be protected in the long term by using symmetric algorithms with long key lengths. However, such a choice would have a severe impact on the storage requirements of the designed blockchain systems. Unless fully homomorphic encryption or some form of malleable encryption schemes is used, the processing of these data will also be impossible. But even then, the extra burden of processing and querying encrypted data would have a severe impact on the performance of the blockchain system \cite{vo2018research}.

Blockchain pruning is proposed as a way to remove data from blockchains. In blockchain pruning, old transactions and blocks are deleted after a predefined amount of time, whereas old block headers containing the hashed version of the removed block data are maintained to ensure the integrity and security of the blockchain. While originally pruning aimed at compressing the blockchain size on the assumption that historical data are not required, it is argued that it can also offer an increased level of user privacy since old transactions might not be locatable. Accordingly, it can serve regulatory requirements allowing the old transactions to be forgotten from the network \cite{farshid2019design,finck2018blockchains,geraud2017twisting}. In this respect, a cryptocurrency scheme called the ``mini-blockchain'' has been proposed as a pruning alternative to current blockchain implementations \cite{brucemini}. The proposed scheme eliminates the need for a full blockchain by unlinking transactions, and therefore it allows all transactions to be discarded after a safe amount of time has elapsed. Obviously, when nodes discard the old blocks, they do not discard the block headers which are stored in a separate ``proof-chain'' to maintain the long term blockchain history. Although blockchain pruning meets scalability and privacy requirements, it has been argued that it does so at the expense of the security since, even when old block headers are maintained, truncating blockchain's history yields to a decreased security \cite{finck2018blockchains}. Pruning has also been criticised for its weak enforceability as there is not any guarantee that all nodes will choose not to store the full chain. Nonetheless, it has been foreseen that pruning may be an appropriate solution for permissioned blockchain frameworks where the operating environment is more easily controlled and adjusted \cite{palm2017implications}. Yet, the idea of pruning in public blockchains remains controversial, and it is nowadays an active field of research \cite{ethereumpruning}.

\subsection{Removing blockchain's immutability}
Much has been written on the advantages and disadvantages of having a mutable blockchain, i.e. a blockchain whose content can be edited or deleted. While for crypto proponents the idea seems repulsive as it eradicates the blockchain's append-only and censorship-resistance nature, for business technocrats the idea seems rather reasonable as it may adapt blockchains to enterprises' requirements and constraints. Despite the arguments on both sides of the debate, the technical implementation for introducing mutability to blockchains is not an easy task. Technologically speaking, the research on removing blockchain's immutability while preserving security is still in infant stages. Yet, some interesting cryptographic and innovative proposals towards this end are discussed below.

Reversing transactions in fraudulent or exceptional cases was discussed among bitcoin developers and blockchain thinkers even from the early days of cryptocurrency boom \cite{bitcoinreverse}. However, since bitcoin was built by design as being immutable for security purposes, crypto supporters were not in favour of such an option. Reversecoin however, was the first altcoin that attempted to reverse transactions within a timeout period \cite{reversecoin}. Its idea was to enable users to seamlessly transact with their online wallets and fall back to an offline wallet if their online account gets hacked \cite{reversecoin2}. Reversecoin worked by setting two different kinds of accounts: Standard Accounts, which are like bitcoin accounts; and Vault Accounts, which are like bank savings accounts. Each vault account has a configurable timeout and is backed by two key pairs, one online and one offline. Only the online key pair is needed to transfer coins from a vault, and the resulting transactions are confirmed after they live in blockchain for the timeout period. During this period, one can reverse those transactions by using the offline key pair and restore the coins in case the transaction originated by a malicious user. Additionally, all reverted transactions remained untouched in the blockchain history so can be publicly viewed. Unfortunately, although reversecoin's original idea was rather appealing, the project did not enjoy widespread acceptance.

The first technical proposal that actually challenged blockchain's immutability is the one published by Ateniese et al. \cite{ateniese2017redactable}. The authors proposed the replacement of the hash function that connects each block to the previous one with an evolution of the standard chameleon hash. A chameleon hash is a cryptographic hash function that contains a trapdoor, and the knowledge of this trapdoor allows collisions to be generated efficiently \cite{KrawczykR00}. While in a standard chameleon hash collisions must be kept private since the trapdoor can be extracted from a single collision, in the proposed improved design it is safe to reveal any number of collisions. With the knowledge of the trapdoor key, it is possible to efficiently find collisions and thus replace the content of the blocks. Thereby, knowing the key, any redaction of the blockchain is possible, including deletion, modification, and insertion of any number of blocks. The proposed system also leaves an immutable ``scar'' to indicate when any blocks have been altered, maintaining thus auditability and transparency. Researchers' main idea was to have the trapdoor key be secretly shared among some fixed set of users that are in charge of redacting the blockchain content in specific and exceptional circumstances. For example, the key could be in the hands of miners, a centralised auditor, or shares of the key could be distributed among several authorities, so that unanimous agreement must be reached to make any changes.

Unavoidably, the announcement of the first redactable blockchain was met with widespread derision while provoked a lot of agitation and scepticism among blockchain believers and cryptocurrency advocates who even argued that an editable blockchain is actually similar to a database \cite{accentureplan,downsideofbitcoin,accenturebreaks,editblockchaindatabase}. They were further claiming that having to trust a set of specific participating authorities, such as banks, to edit the blockchain contents invalidates the decentralised nature of blockchains and defeats the very benefit of this technology \cite{finck2018blockchains,greenspan2017blockchain}. In addition, they argued that a redactable blockchain opens up the financial systems to possible fraudulent activities because the disclosure of the trapdoor key makes the blockchain vulnerable to malicious attacks and decreases its security \cite{accenturesecurespatent}. Despite the criticism, the authors teamed up with Accenture, a big consulting firm, to develop a prototype adapted and refined for permissioned environments based on Hyperledger. Notwithstanding the author's argument that the solution is compatible with current blockchain frameworks, both permissionless and permissioned, sharing the key needed to edit a blockchain to a finite number of trusted nodes renders the solution suitable only for permissioned settings. However, as stated in \cite{greenspan2017blockchain}, in permissioned blockchains mutations can be performed much more easily based on a voting process, albeit less optimised in terms of performance.

Another technical solution for forgetting data stored in blockchains is proposed in \cite{puddu2017muchain} where a mutable blockchain that enables the deletion and modification of blockchain content is described. The proposed design leverages the consensus mechanisms of traditional blockchains to vote on alternate versions of blockchain history. It does so through the introduction of mutable transactions which represent transactions sets that contain various possible versions of transactions. In a transaction set, only one of the transactions is specified as active, while all the others are inactive alternatives. All modifications are performed using transactions of a special type, meta-transactions, which are issued by users or smart contracts and are verified by validators. Mutations are also subject to access control policies specified by the transaction senders. These policies define who, and under which circumstances, is allowed to trigger mutations or to add additional versions of data records, and validators verify their conditions. To hide alternative history versions, the blockchain relies on encryption: all possible transaction versions are encrypted using transaction-specific keys whereas only the decryption keys for the active records are made available. To adapt the setting to the constraints of permissionless blockchains, the authors use a secret sharing scheme to split the transaction-specific keys into shares and distribute those shares among the validators, which can only reconstruct the entire key if a sufficient number of shares are collected. However, as the authors state, this scheme adds significant performance overhead and limits the verification enforcement of some transaction properties. Additionally, while the proposed blockchain offers solutions to the patching of vulnerable smart contracts and the elimination of abusive content from blockchains, it also presents some limitations that hinder its wide acceptance as a forgetting mechanism in permissionless settings. For instance, once an active transaction becomes inactive due to mutation, and therefore its decryption key is not served anymore by validators, local copies of keys may remain stored locally by clients. As a result, the reconstruction of an inactive, i.e. ``forgotten'', record is still possible.

Criticising the above proposal for allowing a malicious user in a public blockchain to simply not include a mutation for his transaction, or even to set a policy where only he himself can mutate the transaction, the authors of \cite{deuberredactable} present a redactable blockchain that does not rely on heavy cryptographic tools and is suitable for permissionless settings. Its protocol uses a consensus-based voting based on a PoW and is parameterized by a policy that dictates the requirements and constraints for the redactions. Any user can propose the edit operations but they are only performed if approved by the blockchain policy (e.g., voted by the majority). Moreover, the protocol offers accountability for edit operations as any edit in the chain can be publicly verified. Nonetheless, although the proof-of-concept implementation of the proposed scheme presents only a tiny overhead in the chain validation when compared to an immutable one, the proposed permissionless blockchain operates on the assumption that the majority of the miners in the network are honest, and they behave rationally when they vote to either accept or reject the edit requests.

In \cite{dorri2019mof} a memory flexible blockchain framework tailored towards IoT networks is presented. The framework allows users to modify, compress, or completely remove their transactions from blockchains while it preserves transactions' consistency. This is achieved by computing the hash of the block over the hashes of its constituted transactions and not of their contents, thereby allowing a transaction to be removed from a block without impacting the hash consistency checks. In particular, for each transaction stored in the blockchain, a specific value is calculated as the signed hash of a secret only the entity generating the transaction knows. To remove a stored transaction, the user has to prove that it has previously generated that transaction by including in the remove transaction the hashes used to generate the secret of the transaction to be removed and the encrypted form of the hashed secret using her public key. When a transaction is removed, while its content is removed from the blockchain, the hash of its content and the hash of its preceding transaction remain stored in the blockchain to ensure blockchain consistency and auditability. To facilitate the removal process, multiple agents are introduced to reduce the packet and processing overhead associated with multiple memory optimisation methods used. Each agent is identified by a unique public key which is certified by a Certificate Authority (CA) to verify its identity. Moreover, for maintaining consistency among transactions and for auditing purposes, a shared read-only central database known as a blackboard and managed centrally by a Blackboard Manager Agent (BMA) is employed. Multiple replications of the blackboard exist to reduce the risk of single point of failure and to ensure scalability. Overall, the proposed framework provides a solid technical framework suitable for compressing, modifying and removing transaction data from blockchains in IoT environments. Yet, since it relies heavily on centralised entities (CA and BMA) for the management of its key functionalities (agents and blackboard), it significantly deviates from a fully decentralised solution.

In another research, the problem of preserving hash-based integrity when deleting transactions from blockchains is tackled \cite{kuhn2018data}. The author describes a data structure, a block matrix, and an algorithm that allow the safe deletion of arbitrary records while preserving hash-based integrity assurance that other blocks remain unchanged. However, the solution has been thus far focused only to permissioned blockchains to ensure their transaction integrity and their compliance with the erasing requirements of the RtbF \cite{supportinggdprreq}. Nevertheless, the idea appears rather appealing as it delves into a core blockchain element, its data structure.

Similarly to blockchain transaction data and contrary to traditional distributed applications that can be patched when bugs are detected, smart contracts living on the blockchain are also irreversible and immutable \cite{luu2016making}. In other words, once smart contracts' code is migrated to the blockchain network there is no way to patch bugs or alter their functionalities. Smart contracts are not removed from the blockchain when their use has come to an end. Instead, they are part of the history of the blockchain and probably retained by most nodes. Even when developers think in advance a way to disable them manually, by inserting ad-hoc code in the contracts, or automatically, by calling self-destruct or suicide functions, the smart contracts are still present but unresponsive \cite{bartoletti2017empirical, introsmartcontracts}. Yet, smart contracts' immutability refers only to their actual code and not to their state which is mostly set from the state of their variables and functions. In fact, in ethereum network, while variables' state can change freely, the history of storage variables in contracts is permanently stored. Furthermore, the functions in the contracts' code are immutable once they are deployed to the blockchain. This immutability is exploited by decentralized applications (DApps) to store some data persistently, and in some cases to certify data ownership and provenance, e.g. to write the hash of a document on the blockchain so that they can prove document existence and integrity \cite{bartoletti2017empirical}. However, due to their immutable nature of smart contracts, their correctness has been identified as a critical factor for their proper and safe behaviour \cite{luu2016making,bhargavan2016formal}. Furthermore, acknowledging that, in contrast to their analogue counterparts, smart contracts' immutability does not allow traditional tools of contract law for termination, rescission, modification and reformation, to be applied successfully to smart contracts, researchers are arguing for a new set of standards to alter and undo smart contracts in order to ensure that the traditional tools achieve their original (contract law) goals when applied to the blockchain technology \cite{marino2016setting}.

\section{Discussion and conclusions }
The controversy over the immutability of blockchain protocols has been given considerable prominence recently due to the adoption of the GDPR and, most importantly, due to the RtbF which foresees the retroactive erasure of personal data upon request and from all available places to which they have been disseminated. Immutability, on the other hand, is fundamental to blockchain's security as it forbids tampering with blockchain data and therefore it facilitates the single, globally accepted view of events among non-trusted participants. In other words, immutability supports the possibility of decentralized trust in inherently trustless interactions. For cryptocurrency activists and blockchain proponents even simply questioning the immutable nature of blockchain is tantamount to heresy \cite{downsideofbitcoin} and therefore they regard the RtbF as an obstacle to the widespread adoption of blockchain technology. On the opposite side, privacy advocates look upon blockchains' immutability as a risk to people's data protection and privacy rights. For enterprise technocrats, however, incorporating limited mutability within permissioned blockchain systems, subject to certain conditions, can strike the right balance between preserving blockchain's key features and adapting it for real-world requirements \cite{accentureunveiledit}. In this perspective, the recent advancements on introducing mutability, based on strict, pre-approved rules, appeals both to regulators and to enterprises \cite{sater2017blockchain}.

In view of the above, the number of public authorities that have started exploring the use of blockchain for their administration and services is rising \cite{olnes2018blockchain,guardtimesecures,canadatrialing}. In 2017, DG TAXUD, the EU General Directorate responsible for EU policies on taxation and customs, started exploring blockchain technology within the customs domain \cite{dgtaxudblockchain}, while a year ago 21 EU Member States plus Norway agreed to sign a declaration creating the European Blockchain Partnership (EBP) and to cooperate in the establishment of a European Blockchain Services Infrastructure (EBSI) that will support the delivery of cross-border digital public services \cite{europeancountries}. At about the same time, the European Commission with the support of the European Parliament launched the EU Blockchain Observatory and Forum with the purpose to encourage governments, industry and citizens to benefit from blockchain opportunities \cite{EUBlockchainObservatory}. Similarly, the OECD has begun investigating the benefits and risks of blockchain for economies and societies \cite{oecdblockchain1,oecdblockchain2}, while the UN is gradually embracing blockchain technology \cite{unusageblockchain}. In the banking sector, the use of digital currencies based on blockchain technology is progressing rapidly as many major banks have already announced blockchain projects to build new digital currencies \cite{centralbank,spanishbanks,chinacentralbank}. Enterprises also consistently engage and invest in the blockchain technology \cite{20companies,mainstreamcompanies,deloittecompaniessurvey,airfranceklm}. Notwithstanding these global initiatives towards a blockchain-enabled era, the blockchain's mass-market adoption is not expected any time soon \cite{nomorehype}. In particular, experts believe that blockchain is now where the web was in 1994 \cite{hyperledger3yearslater}. Indeed, while according to Gartner blockchain is one of the emerged trends in 2018, it is expected to reach a healthy, stable plateau at least in five to 10 years \cite{gartner2018trends}.

In spite of blockchain's slow integration into real-life applications, the extent to which blockchain's incompatibility with data protection and privacy rights occupies scientists and businesses is remarkable. In that respect, and towards researching methods and techniques to accomplish compliance of blockchain protocols with the RtbF, several technical solutions have been put forward. The proposed solutions comprise technical methods broadly used nowadays to bypass the blockchain's collision with the RtbF, as well as cryptographic and other advanced methods aiming at conditionally removing the immutability of the blockchain. In this paper, we attempted, on the one hand, to summarise all these innovative methods and the state-of-the-art techniques and, on the other hand, to provide a comprehensive review of their benefits and limitations when applied in the wild to either permissioned or permissionless blockchain frameworks. In this regard, it is our firm belief that this work will be proved valuable both to industry and to academia.

\section*{Acknowledgments}
This work was supported by the European Commission under the Horizon 2020 Programme (H2020), as part of the projects CyberSec4Europe (\url{https://www.cybersec4europe.eu}) (Grant Agreement no. 830929) and \textit{LOCARD} (\url{https://www.locard.eu}) (Grant Agreement no. 832735).

The content of this article does not reflect the official opinion of the European Union. Responsibility for the information and views expressed herein lies entirely with the authors.


\end{document}